\documentclass{article}

\usepackage[english]{babel}
\usepackage[T1]{fontenc}

\usepackage[letterpaper,top=2cm,bottom=2cm,left=3cm,right=3cm,marginparwidth=1.75cm]{geometry}

\usepackage{amsmath}
\usepackage{graphicx}
\usepackage{url}

\usepackage{authblk}
\author[1]{Drew Ehrlich}
\author[1]{Milad Hakimshafaei}
\author[1]{Oskar Elek}
\affil[1]{University of California, Santa Cruz}

\title{Scaffolding Generation using a 3D Physarum Polycephalum Simulation}

\begin{document}
\maketitle

\begin{abstract}
In this demo, we present a novel technique for approximating topologically optimal scaffoldings for 3D printed objects using a Monte Carlo algorithm based on the foraging behavior of the Physarum polycephalum slime mold. As a case study, we have created a biologically inspired bicycle helmet using this technique that is designed to be effective in resisting impacts. We have created a prototype of this helmet and propose further studies that measure the effectiveness and validity of the design.
\end{abstract}

\maketitle

\section{Introduction}
In 3D printing, scaffolding structures are large factors in determining the efficiency of the use of printing material. For example, the optimization of both interior and exterior scaffolding structures has been extensively studied in optimizing infill \cite{Wu2018}, supports \cite{Tricard2020}, or surface topology \cite{Bian2018}. When specifically considering the optimization of surface topology, a large focus has been on parametric algorithms which take a set of points and optimize the surface. Often these are based on fundamental geometric tessellations like Voronoi, Delaunay, gyroid, and their higher-dimensional extensions  \cite{deBerg2008}.

\begin{figure}
\centering
\includegraphics[width=\columnwidth]{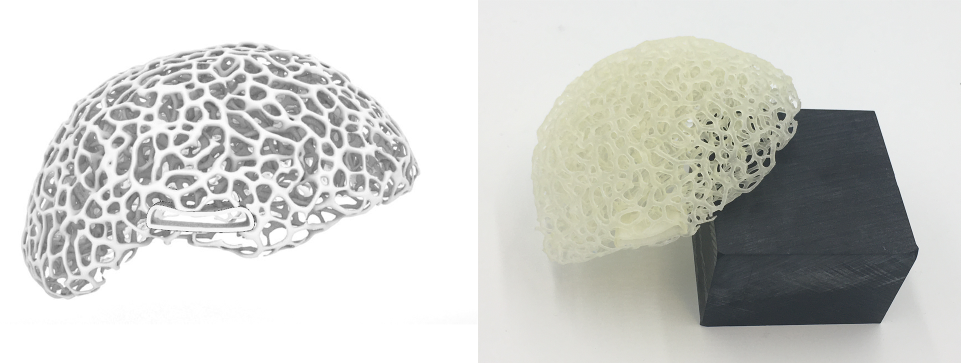}
\caption{The resulting bicycle helmet mesh from our pipeline (left) and a 3D printed version of it created using Siraya Tech Tenacious Resin on a Anycubic Photon Mono SLA 3D printer (right).}
\end{figure}

This work proposes an unconventional methodology for approximating topologically optimal scaffoldings based on the Monte Carlo Physarum Machine algorithm (MCPM,  \cite{Elek_2022}). The MCPM algorithm generalizes the growth and foraging patterns of Physarum Polycephalum slime mold to 3D. These slime mold foraging patterns have been previously shown to have near optimal to optimal geometric properties in 2D and 2.5D contexts  \cite{Adamantzky, Jones2010}. This technique was initially used for analyzing cosmological datasets relating to the `cosmic web', which is a massive transport network that allows for gas to travel through space. The connections created by these simulations have also been previously adapted to create biologically inspired 3D printable models. These models can be generated from generative algorithms, existing voxel grids, or meshes  \cite{ehrlich}. As a case study of the application of the MCPM for additive manufacturing, we are presenting the design and fabrication of a 3D printable bio-inspired bicycle helmet that can be easily adapted to be shaped to the wearer's head with the use of photogrammetry while also providing effective impact protection.

\section{Methods}
To produce our custom fitted bicycle helmet, we designed a pipeline consisting of four components: scanning, modeling, simulating, and reconstructing. We wanted to design the helmet to fit an experienced biker (and co-author), and as such we did a photogrammetric scan of his head. This scan was then converted into a 3D model and voxelized before being imported to Rhinoceros to clean it and remove unnecessary areas from the phtogrammetric scan. To create a simplified version of the model, we optimized the number of polygons generated by the photogrammetric scan by adding thickness to the surface of the model. Once we had a result resembling a bicycle helmet shaped for our biker's head, the resulting set of points from the 3D model was run through PolyPhy, a currently in development open source successor to Polyphorm  \cite{Elek_2020}. Polyphorm and PolyPhy are simulation software that implement the MCPM in a 3D space to create a density field that connects all of the vertices of the imported model into a transport network. This network was then exported as a multidimensional numpy array which was passed through the Marching Cubes \cite{marchingcubes} mesh reconstruction algorithm to create a watertight mesh which was then 3D printed [Figure 1].

\begin{figure}
\centering
\includegraphics[width=\columnwidth]{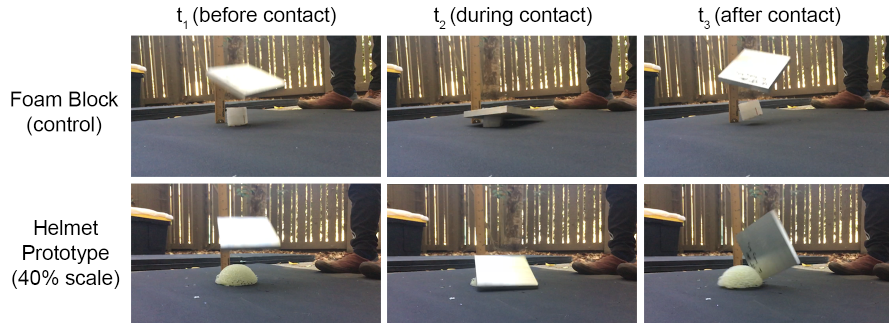}
\caption{The results of a material and structure stress test which demonstrate that the helmet structure printed in Siraya Tech Tenacious resin resists impact forces in a similar manner to a block of foam.}
\end{figure}

\section{Results}
While this work is still in its early stages, we have created objects and gathered data that we would like to receive feedback on from generative design and material science experts in the SCF community. 

We have manufactured a pair of helmets that are around 40\% scale, which were created at a smaller size due to constraints on 3D printers we currently have access to. We are using this as a starting point for testing design efficacy. One of these helmets can be seen in Figure 1 and is printed with Siraya Tech Tenacious Resin \cite{sirayatech}, which is known for its elastomeric properties. This helmet was printed on an Anycubic Photon Mono \cite{anycubic}, a consumer grade SLA 3D printer. We chose this material for its ability to bend without breaking, since most common SLA photosensitive resins will snap or shatter when presented with a strong impact force. We tested this flexibility by taking a 5/8" thick 6061 Aluminum plate and dropping it from 3 feet up on both our helmet and a block of foam of similar height. Using this simple test design, we are able to indicate that our 3D printed bicycle helmet resists strong impacts by bending and bouncing the same way the block of foam did [Figure 2]. This kind of flexibility is desirable in the inner shell of a bicycle helmet to help negate blunt force trauma to the skull and brain from a bicycle crash. Both the choice of dropped object and control were made due to material availability, and we plan to do more through physical quantitative testing and virtual finite element analysis in the future.

\begin{figure}
\centering
\includegraphics[width=\columnwidth]{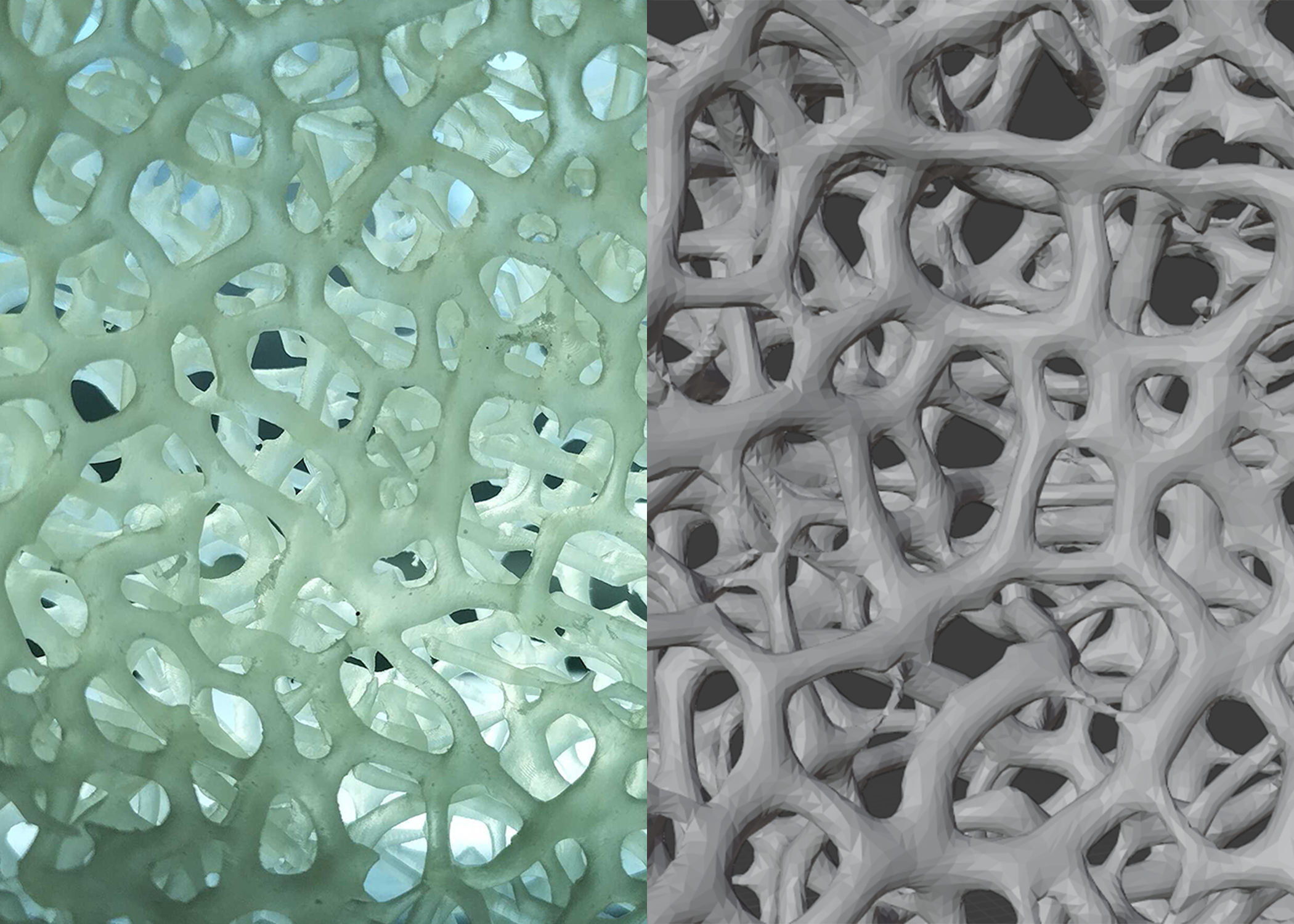}
\caption{A comparison of the details of the resulting 3D printed bike helmet model (left) and the digital mesh that was generated by the MCPM (right).}
\end{figure}

\section{Future Work}

While our current testing scheme is lacking the refinement of the standardized testing used in industrial manufacturing, our experimentation still suggests that our design has some structural merit and is worth exploring further. We will be presenting a full size helmet that is currently being manufactured by a third party service, which will then be part of a series of more rigorous tests after the conference. These tests will range from replicating the tests used for testing bicycle helmets in the USA before going to market \cite{bicyclehelmetstandards} and comparing these results with existing bicycle helmets to strapping the bicycle helmet onto a fragile round object acting as a human head substitute (like a watermelon) to do simple impact tests to see how well the helmet could protect the human head in a real life scenario. 

At SCF, we intend to present plans for these tests along with physical helmet models and possible future applications for the MCPM in the optimization of surface topology. Using the feedback we receive at SCF, we intend to turn our MCPM-based generative design technique into a technical paper doing a more formal analysis of its efficacy and practical uses in additive manufacturing.

\enlargethispage{10pt}

\bibliographystyle{ACM-Reference-Format}
\bibliography{bibl}


\begin{thebibliography}{13}

%%% the Web address.

\ifx \showCODEN    \undefined \def \showCODEN     #1{\unskip}     \fi
\ifx \showDOI      \undefined \def \showDOI       #1{#1}\fi
\ifx \showISBNx    \undefined \def \showISBNx     #1{\unskip}     \fi
\ifx \showISBNxiii \undefined \def \showISBNxiii  #1{\unskip}     \fi
\ifx \showISSN     \undefined \def \showISSN      #1{\unskip}     \fi
\ifx \showLCCN     \undefined \def \showLCCN      #1{\unskip}     \fi
\ifx \shownote     \undefined \def \shownote      #1{#1}          \fi
\ifx \showarticletitle \undefined \def \showarticletitle #1{#1}   \fi
\ifx \showURL      \undefined \def \showURL       {\relax}        \fi
% invisible to TeX
\providecommand\bibfield[2]{#2}
\providecommand\bibinfo[2]{#2}
\providecommand\natexlab[1]{#1}
\providecommand\showeprint[2][]{arXiv:#2}

\bibitem[Adamatzky(2010)]%
        {Adamantzky}
\bibfield{author}{\bibinfo{person}{Andrew Adamatzky}.}
  \bibinfo{year}{2010}\natexlab{}.
\newblock \bibinfo{booktitle}{\emph{Physarum Machines}}.
\newblock \bibinfo{publisher}{WORLD SCIENTIFIC}.
\newblock
\urldef\tempurl%
\url{https://doi.org/10.1142/7968}
\showDOI{\tempurl}
\showeprint{https://www.worldscientific.com/doi/pdf/10.1142/7968}


\bibitem[Anycubic(2022)]%
        {anycubic}
\bibfield{author}{\bibinfo{person}{Anycubic}.} \bibinfo{year}{2022}\natexlab{}.
\newblock \bibinfo{title}{Photon Mono}.
\newblock
\newblock
\urldef\tempurl%
\url{https://www.anycubic.com/products/photon-mono-resin-3d-printer}
\showURL{%
\tempurl}


\bibitem[Bian et~al\mbox{.}(2018)]%
        {Bian2018}
\bibfield{author}{\bibinfo{person}{Xiaojun Bian}, \bibinfo{person}{Li-Yi Wei},
  {and} \bibinfo{person}{Sylvain Lefebvre}.} \bibinfo{year}{2018}\natexlab{}.
\newblock \showarticletitle{Tile-Based Pattern Design with Topology Control}.
\newblock \bibinfo{journal}{\emph{Proc. ACM Comput. Graph. Interact. Tech.}}
  \bibinfo{volume}{1}, \bibinfo{number}{1}, Article \bibinfo{articleno}{23}
  (\bibinfo{date}{jul} \bibinfo{year}{2018}), \bibinfo{numpages}{15}~pages.
\newblock
\urldef\tempurl%
\url{https://doi.org/10.1145/3203204}
\showDOI{\tempurl}


\bibitem[de~Berg et~al\mbox{.}(2008)]%
        {deBerg2008}
\bibfield{author}{\bibinfo{person}{Mark de Berg}, \bibinfo{person}{Otfried
  Cheong}, \bibinfo{person}{Marc van Kreveld}, {and} \bibinfo{person}{Mark
  Overmars}.} \bibinfo{year}{2008}\natexlab{}.
\newblock \bibinfo{booktitle}{\emph{Computational Geometry}}.
\newblock \bibinfo{publisher}{Springer}.
\newblock
\newblock
\shownote{3rd Edition}.


\bibitem[Ehrlich(2021)]%
        {ehrlich}
\bibfield{author}{\bibinfo{person}{Drew Ehrlich}.}
  \bibinfo{year}{2021}\natexlab{}.
\newblock \showarticletitle{Printing the Polyphorm: Using 3D Printing to
  Manufacture Biologically Inspired Rhizomatic Structures}.
\newblock  (\bibinfo{year}{2021}).
\newblock
\urldef\tempurl%
\url{https://drive.google.com/file/d/1-tEPUcKDrB7evDqNNObsLixDUb2R90ZE/view?usp=sharing}
\showURL{%
\tempurl}


\bibitem[Elek et~al\mbox{.}(2020)]%
        {Elek_2020}
\bibfield{author}{\bibinfo{person}{Oskar Elek}, \bibinfo{person}{Joseph~N.
  Burchett}, \bibinfo{person}{J.~Xavier Prochaska}, {and}
  \bibinfo{person}{Angus~G. Forbes}.} \bibinfo{year}{2020}\natexlab{}.
\newblock \bibinfo{title}{Polyphorm: Structural Analysis of Cosmological
  Datasets via Interactive Physarum Polycephalum Visualization}.
\newblock
\newblock
\urldef\tempurl%
\url{https://doi.org/10.48550/ARXIV.2009.02441}
\showDOI{\tempurl}


\bibitem[Elek et~al\mbox{.}(2022)]%
        {Elek_2022}
\bibfield{author}{\bibinfo{person}{Oskar Elek}, \bibinfo{person}{Joseph~N.
  Burchett}, \bibinfo{person}{J.~Xavier Prochaska}, {and}
  \bibinfo{person}{Angus~G. Forbes}.} \bibinfo{year}{2022}\natexlab{}.
\newblock \showarticletitle{Monte Carlo Physarum Machine: Characteristics of
  Pattern Formation in Continuous Stochastic Transport Networks}.
\newblock \bibinfo{journal}{\emph{Artificial Life}} \bibinfo{volume}{28},
  \bibinfo{number}{1} (\bibinfo{year}{2022}), \bibinfo{pages}{22--57}.
\newblock
\urldef\tempurl%
\url{https://doi.org/10.1162/artl_a_00351}
\showDOI{\tempurl}


\bibitem[Institute(2022)]%
        {bicyclehelmetstandards}
\bibfield{author}{\bibinfo{person}{Bicycle Helmet~Safety Institute}.}
  \bibinfo{year}{2022}\natexlab{}.
\newblock \bibinfo{title}{Standards}.
\newblock
\newblock
\urldef\tempurl%
\url{https://helmets.org/#standards}
\showURL{%
\tempurl}


\bibitem[Jones(2010)]%
        {Jones2010}
\bibfield{author}{\bibinfo{person}{Jeff~Dale Jones}.}
  \bibinfo{year}{2010}\natexlab{}.
\newblock \showarticletitle{Characteristics of Pattern Formation and Evolution
  in Approximations of Physarum Transport Networks}.
\newblock \bibinfo{journal}{\emph{Artificial Life}}  \bibinfo{volume}{16}
  (\bibinfo{year}{2010}), \bibinfo{pages}{127--153}.
\newblock


\bibitem[Lorensen and Cline(1987)]%
        {marchingcubes}
\bibfield{author}{\bibinfo{person}{William~E. Lorensen} {and}
  \bibinfo{person}{Harvey~E. Cline}.} \bibinfo{year}{1987}\natexlab{}.
\newblock \showarticletitle{Marching Cubes: A High Resolution 3D Surface
  Construction Algorithm}. In \bibinfo{booktitle}{\emph{Proceedings of the 14th
  Annual Conference on Computer Graphics and Interactive Techniques}}
  \emph{(\bibinfo{series}{SIGGRAPH '87})}. \bibinfo{publisher}{Association for
  Computing Machinery}, \bibinfo{address}{New York, NY, USA},
  \bibinfo{pages}{163–169}.
\newblock
\showISBNx{0897912276}
\urldef\tempurl%
\url{https://doi.org/10.1145/37401.37422}
\showDOI{\tempurl}


\bibitem[Tech(2022)]%
        {sirayatech}
\bibfield{author}{\bibinfo{person}{Siraya Tech}.}
  \bibinfo{year}{2022}\natexlab{}.
\newblock \bibinfo{title}{Siraya Tech Tenacious Resin}.
\newblock
\newblock
\urldef\tempurl%
\url{https://siraya.tech/products/tenacious-resin-flexible-resin}
\showURL{%
\tempurl}


\bibitem[Tricard et~al\mbox{.}(2020)]%
        {Tricard2020}
\bibfield{author}{\bibinfo{person}{Thibault Tricard},
  \bibinfo{person}{Frédéric Claux}, {and} \bibinfo{person}{Sylvain
  Lefebvre}.} \bibinfo{year}{2020}\natexlab{}.
\newblock \showarticletitle{Ribbed Support Vaults for 3D Printing of Hollowed
  Objects}.
\newblock \bibinfo{journal}{\emph{Computer Graphics Forum}}
  \bibinfo{volume}{39}, \bibinfo{number}{1} (\bibinfo{year}{2020}),
  \bibinfo{pages}{147--159}.
\newblock
\urldef\tempurl%
\url{https://doi.org/10.1111/cgf.13750}
\showDOI{\tempurl}
\showeprint{https://onlinelibrary.wiley.com/doi/pdf/10.1111/cgf.13750}


\bibitem[Wu et~al\mbox{.}(2018)]%
        {Wu2018}
\bibfield{author}{\bibinfo{person}{Jun Wu}, \bibinfo{person}{Niels Aage},
  \bibinfo{person}{Rüdiger Westermann}, {and} \bibinfo{person}{Ole Sigmund}.}
  \bibinfo{year}{2018}\natexlab{}.
\newblock \showarticletitle{Infill Optimization for Additive
  Manufacturing—Approaching Bone-Like Porous Structures}.
\newblock \bibinfo{journal}{\emph{IEEE Transactions on Visualization and
  Computer Graphics}} \bibinfo{volume}{24}, \bibinfo{number}{2}
  (\bibinfo{year}{2018}), \bibinfo{pages}{1127--1140}.
\newblock
\urldef\tempurl%
\url{https://doi.org/10.1109/TVCG.2017.2655523}
\showDOI{\tempurl}


\end{thebibliography}

\end{document}